# LOCAL AND GLOBAL EXISTENCE OF MULTIPLE WAVES NEAR FORMAL APPROXIMATIONS

XIAO-BIAO LIN

**Introduction**

The formation of multiple wave fronts is important in applications of singularly perturbed parabolic systems. These solutions can be effectively constructed by formal asymptotic methods. When truncated to a certain order in $\epsilon$, they become formal approximations of solutions to the given system. The precision of a formal approximation is judged by the smallness of the residual error in each regular and singular layer and the jump error between adjacent layers.

The purpose of this paper is to introduce *Spatial Shadowing Lemmas* that help to construct exact solutions near the formal approximations. The *Shadowing Lemma* was first developed for discrete mappings in $\mathbb{R}^n$, see [8]. It has been extended to continuous flows governed by ODEs [11], and semiflows governed by abstract parabolic equations [2]. We will call these *Temporal Shadowing Lemmas* since the dynamical systems considered there evolve in time.

The *Temporal Shadowing Lemma* has been used to construct exact solutions for singularly perturbed ODEs [11]. However, it cannot be applied directly to singularly perturbed parabolic equations. For formal approximations of multiple waves, the jumps between adjacent layers are functions of $t$ and they occur along the $x$-direction. Since parabolic equations cannot be solved in the $x$-direction, therefore, they do not define a dynamical system in the spatial direction.

To solve this problem, we use an idea motivated by the works of Kirchässner and Renardy [10, 15]. We find stable and unstable subspaces of the trace space so that the parabolic system can be solved forward and backward in the spatial direction. Thus the jumps along the lateral common boundaries can be corrected using the technique of the usual *Shadowing Lemma* in abstract spaces.

This paper is divided into two parts. In the Part I, we show that for a general parabolic system, if a formal approximation is precise enough, then there is an exact solution near the formal approximation for at least a short time. The result obtained here applies to various systems including reaction-diffusion equations, Cahn-Hilliard equations [1], and viscous profile of conservation laws. In Part II, we show that with additional restrictions, the process in Part I can be repeated

*Date*: January 26, 1996.
Research partially supported by NSFgrant DMS9002803 and DMS9205535.

to obtain global solutions if the formal approximation is a global one. Examples include reaction-diffusion equations and phase field equations [7].

**PART I. Local Existence of Multiple Waves**

**1.** Consider a general singularly perturbed parabolic system,

$$(1) \quad \epsilon u_t + (-\epsilon^2)^m D_x^{2m} u = f(u, \epsilon u_x, \cdots, (\epsilon D_x)^{2m-1} u, x, \epsilon), \quad u \in \mathbb{R}^n,\ x \in \mathbb{R},$$

where $f$ is $C^\infty$ with bounded derivatives in all the variables. Assume that the system has regular and internal layers located alternatively along the $x$ axis. For simplicity, we solve the system for $x \in \mathbb{R}$, with no boundary conditions other than $u \in H^{2m,1}$. Assume that there are curves $\Gamma^i = \{(x,t) : x = x^i(t),\ t \in [0, \Delta t]\}$, $i \in \mathbb{Z}$, that divide the domain $x \in \mathbb{R}$, $t \in [0, \Delta t]$ into regular or singular layers $\Sigma^i$, each is between $\Gamma^{i-1}$ and $\Gamma^i$.

Assume that a formal approximation is given, piecewise continuous, with $u = \tilde{w}^i(x,t,\epsilon)$ in $\Sigma^i$. Using the stretched variables $\xi = x/\epsilon$, $\tau = t/\epsilon$, let $w^i(\xi,\tau,\epsilon) = \tilde{w}^i(\epsilon\xi, \epsilon\tau, \epsilon)$ which depends slowly on $\tau$. Assume that $w^i \in H^{2m,1}(\Sigma^i)$. Let $W^i = (w^i, D_\xi w^i, \cdots, D_\xi^{2m-1})^\tau$. The error terms in the followings are $-g^i$ and $-\delta^i$,

$$(2) \quad w_\tau^i + (-1)^m D_\xi^{2m} w^i - f(w^i, \cdots, \epsilon\xi, \epsilon) = -g^i, \quad \text{in } \Sigma^i,$$

$$(3) \quad W^i(\xi^i, \tau) - W^{i+1}(\xi^i, \tau) = -\delta^i(\tau), \quad \text{at } \Gamma^i.$$

Let an exact solution be $u^i + w^i$ in $\Sigma^i$. Let $U^i = (u^i, D_\xi u^i, \cdots, D_\xi^{2m-1} u^i)^\tau$. Let $\tau \in I = [0, \Delta\tau]$ where $\Delta\tau$ is independent of $\epsilon$. The domain $[0, \Delta\tau]$ corresponds to a short time interval $[0, \epsilon\Delta\tau]$ in the $t$ variable. Assume that $\epsilon$ is small so that a near identity change of coordinates in $\Sigma^i$ can be used to straighten the boundaries $\Gamma^{i-1}$ and $\Gamma^i$. In the following, we assume that $\xi^i = x^i/\epsilon$ is independent of $\tau$, but may depend on $\epsilon$, $\Omega^i = (\xi^{i-1}, \xi^i)$, and $\Sigma^i = \Omega^i \times I$. With the new coordinates introduced above, linearizing (1) around $w^i$ at the fixed time $\tau = 0$, we have

$$(4) \quad u_\tau^i + (-1)^m D_\xi^{2m} u^i - \sum_{j=1}^{2m-1} A_j^i(\xi) D_\xi^j u^i = \mathcal{N}^i(u^i, g^i, \epsilon), \quad \text{in } \Sigma^i,$$

$$(5) \quad U^i(\xi^i, \tau) - U^{i+1}(\xi^i, \tau) = \delta^i(\tau), \quad \text{at } \Gamma^i,$$

$$(6) \quad u^i(\xi, 0) = u_0^i(\xi), \quad \text{in } \Omega^i.$$

Here $A_j^i(\xi) = D_j f(w^i(\xi, 0, \epsilon), D_\xi w^i(\xi, 0, \epsilon), \cdots)$ is the partial derivative of $f$ with respect to the $j$-th variable. $u_0^i \in H^m(\Omega^i)$. $\mathcal{N}^i$ depends slowly on $\tau$ and $|\mathcal{N}^i|_{L^2(\Sigma^i)} = O(|u^i|^2_{H^{2m,1}(\Sigma^i)} + |g^i|_{\Sigma^i} + |\epsilon\Delta\tau||u^i|_{H^{2m,1}(\Sigma^i)})$.

We look for a sequence of solutions $\{u^i\}_{-\infty}^\infty$ with $u^i \in H^{2m,1}(\Sigma^i)$. Let $H^k(I)$ be the usual Sobolev space and let $H_0^k(I)$ be the completion of $C^\infty$ functions, which are zero in a neighborhood of $\tau = 0$, in the $H^k$ norm. Define the product spaces

$$B^m(I) = \Pi_{k=0}^{2m-1} H^{1-\frac{2k+1}{4m}}(I), \quad B_0^m(I) = \Pi_{k=0}^{2m-1} H_0^{1-\frac{2k+1}{4m}}(I).$$

From the Trace Theorem, [14], the mapping
$$\xi \to U(\xi, \cdot), \ \Omega^i \to B^m(I),$$
is continuous with
$$|U^i(\xi, \cdot)|_{B^m(I)} \leq C|u^i|_{H^{2m,1}(\Sigma^i)}.$$

Therefore, $\delta^i \in B^m(I)$ in (3) and (5). Let $U_0^i = (u_0^i, \cdots, D_\xi^{m-1} u_0^i)^\tau$. Let $\pi_j$ be the projection to the first $j$-tuples in the product space $\Pi_{j=1}^{2m} \mathbb{R}^n$, i.e., $\pi_j(u_1, \cdots, u_{2m}) = (u_1, \cdots, u_j)$, A compatibility condition is also assumed on the initial data and the jumps,

(7) $$U_0^i(\xi^i) - U_0^{i+1}(\xi^i) = \pi_m \delta^i(0).$$

Notice that only the first $m$ components of $\delta^i$ have well defined traces at $\tau = 0$.

We now discuss the method of solving (4)–(6) with the compatibility (7) in Sections **2**–**5**.

**2.** Consider
$$u_\tau + (-1)^m D_\xi^{2m} u = 0, \quad \xi \in \mathbb{R}, \ \tau \in \mathbb{R}^+,$$
with $u(\xi, 0) = 0$. Applying the Laplace transform, we have the so called dual system,

(8) $$\hat{U}_\xi = J(s)\hat{U} = \begin{pmatrix} 0 & I & 0 & \cdots & 0 \\ 0 & 0 & I & \cdots & 0 \\ & & & \cdots & \\ (-1)^{m+1}s & 0 & 0 & \cdots & 0 \end{pmatrix} \hat{U}.$$

The matrix $J(s)$ has $2m$ eigenvalues $\lambda = [(-1)^{m+1}s]^{\frac{1}{2m}}$.

Consider a sector in $\mathbb{C}$,
$$\mathcal{S}_\theta(M) = \{s : |s| \geq M, |arg(s)| \leq \theta\}, \quad M > 0, \pi/2 < \theta < \pi.$$

When $s \in \mathcal{S}_\theta(M)$, the eigenvalues $\lambda$ are in $2m$ disjoint sectors of $\mathbb{C}$, with $|\text{Re}\lambda| \geq \cos(\frac{\pi-\theta}{2m}) \sqrt[2m]{|s|}$. There are $m$ eigenvalues with positive real parts and $m$ with negative real parts. Each eigenvalue has an $n$-dimensional eigenspace spanned by $(u, \lambda u, \cdots, \lambda^{2m-1} u)^\tau$, $u \in \mathbb{R}^n$.

Let $E^{m,\nu}(s)$ be the Banach space of points in $\mathbb{R}^{2mn}$ with an $s$-dependent norm,
$$|(u_0, u_1, \cdots, u_{2m-1})|_{E^{m,\nu}(s)} = \sum_{j=0}^{2m-1} (1 + |s|^{\nu + \frac{j}{2m}})|u_{2m-1-j}|_{\mathbb{R}^n}.$$

We actually will only use $\nu = 0$ or $\frac{1}{4m}$ in this paper. Let $P_s$ and $P_u$ be the projections in $\mathbb{R}^{2mn}$ to the stable and unstable spaces of $J(s)$, $s \in \mathcal{S}_\theta(M)$. The projections are of rank $mn$ and can be constructed using eigenvalues and eigenvectors. Using the $E^{m,\nu}(s)$ norm, we can show that there exist $K$, $\alpha_1$, $\alpha > 0$, such

that

$$|e^{J(s)\xi} P_s|_{E^{m,\nu}(s)} \le K e^{-\alpha_1 \sqrt[2m]{|s|}\xi} \le K e^{-\alpha(1+\sqrt[2m]{|s|})\xi}, \quad \xi \ge 0,$$
$$|e^{J(s)\xi} P_u|_{E^{m,\nu}(s)} \le K e^{\alpha_1 \sqrt[2m]{|s|}\xi} \le K e^{\alpha(1+\sqrt[2m]{|s|})\xi}, \quad \xi \le 0. \tag{9}$$

**3.** Consider

$$u_\tau + (-1)^m D_\xi^{2m} u - \sum_{j=1}^{2m-1} A_j^i(\xi) D_\xi^j u = 0, \quad \xi \in [\xi^{i-1}, \xi^i], \tag{10}$$

with $u(\xi, 0) = 0$. Using the Laplace transform, we have a dual system,

$$\hat{U}_\xi = J(s)\hat{U} + (-1)^m \begin{pmatrix} 0 & 0 & \cdots & 0 \\ & & \cdots & \\ A_0^i(\xi) & A_1^i(\xi) & \cdots & A_{2m-1}^i(\xi) \end{pmatrix} \hat{U}. \tag{11}$$

Let $T^i(\xi, \zeta, s)$ be the solution matrix for system (11). System (11) is said to have an exponential dichotomy in $E^{m,\nu}(s)$ for $\xi \in [\xi^{i-1}, \xi^i]$, $s \in \mathcal{S}_\theta(M)$ if there exist projections $P_s^i(\xi, s) + P_u^i(\xi, s) = I$ in $E^{m,\nu}(s)$, continuous in $\xi$ and analytic in $s$, and positive constants $K$, $\alpha$, such that

$$|T^i(\xi, \zeta, s) P_s^i(\zeta, s)|_{E^{m,\nu}(s)} \le K e^{-\alpha(1+\sqrt[2m]{|s|})|\xi-\zeta|}, \quad \xi \ge \zeta,$$
$$|T^i(\xi, \zeta, s) P_u^i(\zeta, s)|_{E^{m,\nu}(s)} \le K e^{-\alpha(1+\sqrt[2m]{|s|})|\xi-\zeta|}, \quad \xi \le \zeta.$$

Suppose that $\sup_{\xi \in [a,b], 0 \le k \le 2m-1} |A_k^i(\xi)| \le C$. From the Roughness of Exponential Dichotomy Theorem, [3], which is also valid in the Banach space $E^{m,\nu}(s)$, we find that there exists $M = M(C) > 0$, sufficiently large, such that (11) has an exponential dichotomy in $E^{m,\nu}(s)$ for $\xi \in [\xi^{i-1}, \xi^i]$ and $s \in \mathcal{S}_\theta(M)$. On the other hand, if $M$ is fixed, then there exists $C = C(M) > 0$, sufficiently small, such that the same conclusion holds.

A function $f(s)$ is in the Hardy-Lebesgue class $\mathcal{H}(\gamma)$, $\gamma \in \mathbb{R}$, if
  (i) $f(s)$ is analytic in $\operatorname{Re}(s) > \gamma$;
  (ii) $\{\sup_{\sigma > \gamma} (\int_{-\infty}^{\infty} |f(\sigma + i\omega)|^2 d\omega)\}^{1/2} < \infty$.
$\mathcal{H}(\gamma)$ is a Banach space with the norm defined by the left side of (ii). Based on the Paley-Wiener Theorem, [16], if $e^{-\gamma t} f(t) \in L^2(\mathbb{R}^+)$, then $\hat{f}(s) \in \mathcal{H}(\gamma)$, vice versa.

For $k \ge 0$ and $\gamma \in \mathbb{R}$, define a Banach space

$$\mathcal{H}^k(\gamma) = \{u(s) \mid u(s) \text{ and } (s-\gamma)^k u(s) \in \mathcal{H}(\gamma)\},$$
$$|u|_{\mathcal{H}^k(\gamma)} = |u|_{\mathcal{H}(\gamma)} + |(s-\gamma)^k u|_{\mathcal{H}(\gamma)}.$$

For any $\gamma \in \mathbb{R}$, $k \ge 0$, there exists $C = C(\gamma, k)$ such that

$$C^{-1}(1 + |s|^k) \le 1 + |s - \gamma|^k \le C(1 + |s|^k).$$

Therefore an equivalent norm for $\mathcal{H}^k(\gamma)$ is

$$|u|^2_{\mathcal{H}^k(\gamma)} = \sup_{\sigma > \gamma} \int_{-\infty}^{\infty} |u(\sigma + i\omega)|^2 (1 + |\sigma + i\omega|^{2k}) d\omega.$$

It can be shown that if $e^{-\gamma t} f(t) \in H_0^k(\mathbb{R}^+)$, then $\hat{f}(s) \in \mathcal{H}^k(\gamma)$.

We often use Banach spaces of functions with norms weighted by $e^{-\gamma\tau}$. For example $B_0^m(\mathbb{R}^+,\gamma) = \{\phi : e^{-\gamma\tau}\phi(\tau) \in B_0^m(\mathbb{R}+)\}$ with obvious norms. Other spaces like $L^2(\mathbb{R} \times \mathbb{R}^+,\gamma)$, etc. can be defined similarly.

Let $\mathcal{K}^m(\gamma) = \Pi_{k=0}^{2m-1}\mathcal{H}^{1-\frac{2k+1}{4m}}(\gamma)$. From the definition of $B_0^m(\mathbb{R}^+,\gamma)$, we see that
$$\phi \in B_0^m(\mathbb{R}^+,\gamma) \Leftrightarrow \hat{\phi} \in \mathcal{K}^m(\gamma).$$

Suppose now the system (11) has an exponential dichotomy in $E^{m,\frac{1}{4m}}(s)$ for $\xi \in [\xi^{i-1}\xi^i]$ and $\operatorname{Re} s > \gamma$. An equivalent norm for $\mathcal{K}^m(\gamma)$ is
$$|\phi|_{\mathcal{K}^m(\gamma)} \sim \sup_{\sigma>\gamma}[\int_{-\infty}^{\infty} |\phi(s)|_{E^{m,\frac{1}{4m}}(s)} d\omega]^{1/2}, \quad s = \sigma + i\omega.$$

Based on this, one can show that (11) also has an exponential dichotomy in $\mathcal{K}^m(\gamma)$ for $\xi \in [\xi^{i-1},\xi^i]$. The definition for exponential dichotomies in a Banach space like $\mathcal{K}^m(\gamma)$ is standard, and can be found in [9]. Using the definitions of the function spaces and exponential dichotomies, it is not hard to show the following. (See [13] for the case $m = 1$.)

**Lemma 1.** *For any $\phi \in B_0^m(\mathbb{R}^+,\gamma)$, consider*
$$u_1 = \pi_1 \mathcal{L}^{-1}(T^i(\xi,\xi^{i-1};s)P_s^i(\xi^{i-1},s)\hat{\phi}(s)), \quad \xi \geq \xi^{i-1},$$
$$u_2 = \pi_1 \mathcal{L}^{-1}(T^i(\xi,\xi^i;s)P_u^i(\xi^i,s)\hat{\phi}(s)), \quad \xi \leq \xi^i.$$

*If $\sup_{\xi\in[\xi^{i-1},\xi^i],0\leq k\leq 2m-1}|A_k(\xi)| < \infty$, then $u_j \in H^{2m,1}([\xi^{i-1},\xi^i] \times \mathbb{R}^+,\gamma)$, $j = 1,2$ and is a solution to (10) with $u_j(\xi,0) = 0$. Moreover*
$$|u_j|_{H^{2m,1}(\gamma)} \leq C|\phi|_{B_0^m(\mathbb{R}^+,\gamma)}.$$

**4.** A sequence of functions $f^i \in F^i$, $i \in \mathbb{Z}$, where $F^i$ is a Banach space, will be denoted by $\{f^i\}$. Define the norm $|\{f^i\}|_{F^i} = \sup_i\{|f^i|_{F^i}\}$.

Consider a linear system,
$$(12) \quad u_\tau^i + (-1)^m D_\xi^{2m} u^i - \sum_{j=1}^{2m-1} A_j^i(\xi) D_\xi^j u^i = h^i(\xi,\tau), \quad \text{in } \Sigma^i,$$

with jump conditions (5), initial conditions (6) and compatibilities (7). Assume that the coefficients $A_j^i(\xi)$ are extended to $\xi \in \mathbb{R}$ by constants outside $\Omega^i$. After the extension, assume that the associated homogeneous dual system (11) has exponential dichotomies in $E^{m,\nu}(s)$, $\nu = 0, 1/4m$, for $\xi \in \mathbb{R}$ and $\operatorname{Re}(s) > \gamma$. Assume that $g^i \in L^2(\Sigma^i,\gamma)$, $\delta^i \in B^m(\mathbb{R}^+,\gamma)$ and $u_0^i \in H^m(\Omega^i)$. In this section $\Sigma^i = \Omega^i \times I$ with $I = \mathbb{R}^+$. Assume
$$|\{\delta^i\}|_{B^m(\mathbb{R}^+,\gamma)} + |\{h^i\}|_{L^2(\Sigma^i,\gamma)} + |\{u_0^i\}|_{H^m(\Omega^i)} < \infty,$$

We look for solutions $u^i \in H^{2m,1}(\Sigma^i)$, $i \in \mathbb{Z}$.

By a standard method, we can continuously extend $h^i$ and $u_0^i$ to $\xi \in \mathbb{R}$ so that
$$|h^i|_{L^2(\mathbb{R}\times\mathbb{R}^+)} \leq C|h^i|_{L^2(\Sigma^i)}, \quad |u_0^i|_{\in H^m(\mathbb{R})} \leq C|u_0^i|_{H^m(\Omega^i)}.$$

We first solve (12) in the domain $\mathbb{R} \times \mathbb{R}^+$, with an initial condition $u_0^i$ but no jump conditions. Denote the solution by $\bar{u}^i$. From the existence of exponential dichotomy in $E^{m,0}(s)$, for $\text{Re}(s) > \gamma$, we can prove that (12) defines a sectorial operator $\mathcal{A}^i$ in $L^2(\mathbb{R})$. Moreover,

$$|\lambda - \mathcal{A}^i|_{L^2(\mathbb{R})}^{-1} \leq \frac{C}{1+|\lambda|}, \quad \text{Re}\lambda > \gamma.$$

The proof of the case $m = 1$ can be found in [13]. The general case can be proved similarly. It is than easy to see that $\bar{u}^i$ can be solved by the analytic semigroup $e^{\mathcal{A}^i \tau}$ and the variation of constant formula. We have

$$|\bar{u}^i|_{H^{2m,1}(\mathbb{R}\times\mathbb{R}^+,\gamma)} \leq C(|h^i|_{L^2(\mathbb{R}\times\mathbb{R}^+,\gamma)} + |u_0^i|_{H^m}).$$

Let $\bar{U}^i = (\bar{u}^i, D_\xi \bar{u}^i, \cdots, D_\xi^{2m-1}\bar{u}^i)^\tau$, and $\bar{\delta}^i = \bar{U}^i(\xi^i,\cdot) - \bar{U}^{i+1}(\xi^i,\cdot)$. Then

$$|\bar{\delta}^i|_{B^m(\mathbb{R}^+,\gamma)} \leq C(|\{h^i\}|_{L^2(\Sigma^i,\gamma)} + |\{u_0^i\}|_{H^m(\Omega^i)}).$$

Let the solution to (12), (5)–(7) be $u^i = \bar{u}^i + v^i$. The function $v^i$ satisfies (12) with $h^i = 0$ and $v^i(\xi,0) = 0$. Let $\eta^i = \delta^i - \bar{\delta}^i$. Then $\eta^i \in B_0^m(\mathbb{R}^+, \gamma)$. Let $V^i = (v^i, D_\xi v^i, \cdots, D_\xi^{2m-1}v^i)^\tau$. Then the dual systems for $V^i$ are

$$(13) \qquad \hat{V}_\xi^i = J(s)\hat{V}^i + (-1)^m \begin{pmatrix} 0 & 0 & \cdots & 0 \\ & & \cdots & \\ A_0^i(\xi) & A_1^i(\xi) & \cdots & A_{2m-1}^i(\xi) \end{pmatrix} \hat{V}^i.$$

$$(14) \qquad \hat{V}^i(\xi^i,\cdot) - \hat{V}^{i+1}(\xi^i,\cdot) = \hat{\eta}^i.$$

We want to solve (13) and (14) with $\hat{\eta}^i \in \mathcal{K}^m(\gamma)$.

$$|\hat{\eta}^i|_{\mathcal{K}^m(\gamma)} \leq C(|\delta^i|_{B^m(\mathbb{R}^+,\gamma)} + |\{h^i\}|_{L^2(\Sigma^i,\gamma)} + |\{u_0^i\}|_{H^m}).$$

For two subspaces $M \oplus N = \mathbb{R}^{2mn}$, denote by $P(M,N)$ the projection with the range and kernel being $M$ and $N$ respectively. Assume that at each $\xi^i$, we have

$$\mathcal{R}P_u^i(\xi^i,s) \oplus \mathcal{R}P_s^{i+1}(\xi^i,s) = \mathbb{R}^{2mn}.$$

Here $\mathcal{R}$ stands for the range of an operator, and $P_u^i$ and $P_s^{i+1}$ are projections associated to the exponential dichotomies in $\Omega^i$ and $\Omega^{i+1}$ respectively. Assume that the norms of the projections associated with the above splitting are uniformly bounded with respect to $i \in \mathbb{Z}$, $\text{Re}\,s > \gamma$ in the $E^{m,\nu}(s)$ norm. Notice that the assumption is valid if we choose $\gamma > 0$ to be sufficiently large, due to the Roughness of Exponential Dichotomies again.

We now solve (13), (14) by an iteration method that is used to prove the *Temporal Shadowing Lemma*, [11]. As a first approximation, let

$$\phi_u^i(\xi^i,s) = P(\mathcal{R}P_u^i(\xi^i,s), \mathcal{R}P_s^{i+1}(\xi^i,s))\hat{\eta}^i(s),$$
$$\phi_s^i(\xi^{i-1},s) = -P(\mathcal{R}P_s^i(\xi^{i-1},s), \mathcal{R}P_u^{i-1}(\xi^{i-1},s))\hat{\eta}^{i-1}(s),$$
$$\phi^i(\xi,s) = T^i(\xi,\xi^{i-1},s)\phi_s^i(\xi^{i-1},s) + T^i(\xi,\xi^i,s)\phi_u^i(\xi,s).$$

From Lemma 1, $\pi_1 \mathcal{L}^{-1}(\phi^i(\xi,s)) \in H^{2m,1}(\Omega^i \times \mathbb{R}^+, \gamma)$ and is a solution for (12), with $h^i = 0$. However, at $\xi^i$, the jump is not exactly $\hat{\eta}^i$. Let $\phi^i(\xi^i, \cdot) - \phi^{i+1}(\xi^i, \cdot) = \hat{\eta}^i - \hat{\eta}_1^i$. Then

$$\hat{\eta}_1^i(s) = T^{i+1}(\xi^i, \xi^{i+1}, s)\phi_u^{i+1}(\xi^i, s) - T^i(\xi^i, \xi^{i-1}, s)\phi_s^i(\xi^{i-1}, s).$$

$$|\{\hat{\eta}_1^i\}|_{\mathcal{K}^m(\gamma)} \leq Ce^{-\alpha d}|\{\hat{\eta}^i\}|_{\mathcal{K}^m(\gamma)}.$$

Here $d = \inf\{\xi^{i+1} - \xi^i\}$. The above process can be repeated with $\{\hat{\eta}^i\}$ replaced by $\{\hat{\eta}_1^i\}$, and the second approximation denoted by $\{\phi_1^i\}$. By iteration, we can have a sequence $\{\hat{\eta}_j^i\}$, $j \geq 1$ and a sequence of approximations $\{\phi_j^i\}$. Suppose now the constant $C_1 = Ce^{-\alpha d} < 1$, then the convergent series

$$\Phi^i = \phi^i + \sum_{j=1}^{\infty} \phi_j^i$$

is the desired solution to (13) and (14). $\pi_1 \mathcal{L}^{-1}(\Phi^i)$ is an exact solution for $v^i$. The uniqueness of $\{v^i\}$ can be proved by the exponential dichotomy argument and will be skipped here. Observe that by the Paley-Wiener Theorem,

(15) $\qquad \eta^i(\tau) = 0$ for $\tau \leq \Delta\tau$, $i \in \mathbb{Z}$, $\Rightarrow$ $v^i(\xi, \tau) = 0$ for $\tau \leq \Delta\tau$, $i \in \mathbb{Z}$.

This fact will be used in the next section.

Finally, the solution to the system (12), (5)–(7), $u^i = \bar{u}^i + v^i$, satisfies

(16) $\qquad |u^i|_{H^{2m,1}(\Sigma^i, \gamma)} \leq C(|\{\delta^i\}|_{B^m(\mathbb{R}^+, \gamma)} + |\{h^i\}|_{L^2(\Sigma^i, \gamma)} + |\{u_0^i\}|_{H^m(\Omega^i)}).$

**5.** The nonlinear system (4)–(7) can be solved by using the result of §4 on the linear system and a contraction mapping in a suitable Banach space. The following *Local Spatial Shadowing Lemma* is the main result of Part I.

**Theorem 2.** *Assume that $f$ is $C^\infty$ with bounded derivatives with respect to all the variables, $\{w^i\}$ is a formal approximation with $|\{w^i\}|_{H^{2m,1}(\Sigma^i)} < \infty$. Let $I = [0, \Delta\tau]$. Assume that $\epsilon > 0$ is small so that a near identity change of coordinates can be made in $[0, \epsilon\Delta\tau]$ as in §1. Let $d = \inf_i\{\xi^{i+1} - \xi^i\} > 0$. Then there exist $\beta_0$, $\epsilon_0 > 0$ and a positive linear function $\mu^*(\beta)$, $0 < \beta \leq \beta_0$. If $0 < \epsilon < \epsilon_0$, and*

$$|\{u_0^i\}|_{H^m(\Omega^i)} + |\{g^i\}|_{L^2(\Sigma^i)} + |\{\delta^i\}|_{B^m(I)} \leq \mu^*(\beta),$$

*then there is a unique solution $\{u^i\}$ to (4)–(7), with $|\{u^i\}|_{H^{2m,1}(\Sigma^i)} \leq \beta$. Moreover,*

$$|\{u^i\}|_{H^{2m,1}(\Sigma^i)} \leq C(|\{u_0^i\}|_{H^m(\Omega^i)} + |\{g^i\}|_{L^2(\Sigma^i)} + |\{\delta^i\}|_{B^m(I)}).$$

*Proof.* Let $h^i \in L^2(\Sigma^i)$, $i \in \mathbb{Z}$. Since $\gamma > 0$, it is easy to extend the domains of $\delta^i$ and $h^i$ to $\tau \in \mathbb{R}^+$, so that

$$|\{h^i\}|_{L^2(\Omega^i \times \mathbb{R}^+, \gamma)} + |\{\delta^i\}|_{B^m(\mathbb{R}^+, \gamma)} \leq C(|\{h^i\}|_{L^2(\Sigma^i)} + |\{\delta^i\}|_{B^m(I)}).$$

Consider the associated linear system (12). From the assumptions, it is clear that $\sup_{\xi,i,k} |A_k^i(\xi)| < \infty$. Assume that the coefficients have been extended to $\xi \in \mathbb{R}$ by constants, then from §4, there exists $M_0 > 0$ such that if $M \geq M_0$ then (12) has an exponential dichotomy in $E^{m\nu}(s)$, $\nu = 0, \frac{1}{4m}$ for $\xi \in \mathbb{R}$ and $s \in \mathcal{S}_\theta(M)$.

This also implies that (12) has an exponential dichotomy in $\mathcal{K}^m(\gamma)$ for $\gamma = M$, of which the exponential decay rate is $\tilde{\alpha} = \alpha(1 + \sqrt[2m]{M})$. By choosing larger $M$, we have $C_1 = Ce^{-\tilde{\alpha}d} \leq 0.5$, where $C_1$ is as in §4. The result in §4 concerning the system (12), (5)–(7) is now valid. Let the unique solution be denoted by

$$\{u^i\} = \mathcal{F}_\gamma(\{u_0^i\}, \{\delta^i\}, \{h^i\}).$$

Restricting the solution to the finite interval $I = [0, \Delta\tau]$, in the unweighted norm, using (16), we have,

$$|\{u^i\}|_{H^{2m,1}(\Sigma^i)} \leq Ce^{\gamma\Delta\tau}|\{u^i\}|_{H^{2m,1}(\Omega^i \times \mathbb{R}^+, \gamma)}$$
$$\leq Ce^{\gamma\Delta\tau}(|\{u_0^i\}|_{H^m(\Omega^i)} + |\{h^i\}|_{L^2(\Sigma^i)} + |\{\delta^i\}|_{B^m(I)}).$$

Let the solution in that finite time interval $I$ be denoted by

$$\{u^i\} = \mathcal{F}_I(\{u_0^i\}, \{\delta^i\}, \{h^i\}).$$

Consider $Q(\beta) = \{\{u^i\} : u^i \in H^{2m,1}(\Sigma^i), |\{u^i\}|_{H^{2m,1}} \leq \beta\}$. Let $|\{u_0^i\}|_{H^m(\Omega^i} + |\{\delta^i\}|_{B^m(I)} + |\{g^i\}|_{L^2(\Sigma^i)} = \mu$. For $\{u^i\} \in Q(\beta)$, we have

$$|\mathcal{N}^i(u^i, g^i, \epsilon)|_{L^2(\Sigma^i)} \leq |g^i|_{L^2} + C(|u^i|^2 + \epsilon\Delta\tau|u^i|)$$
$$\leq C(\beta^2 + \epsilon\Delta\tau\beta + \mu).$$

Consider the mapping

$$\{u_1^i\} = \mathcal{F}_I(\{u_0^i\}, \{\delta^i\}, \{\mathcal{N}^i(u^i, g^i, \epsilon)\}).$$

We have

$$|\{u_1^i\}|_{H^{2m,1}(\Sigma^i)} \leq C(\mu + \beta^2 + \epsilon\Delta\tau\beta).$$

Let $\beta$ be small such that $C\beta^2 < \frac{1}{3}\beta$. Let $\mu$ and $\epsilon_0$ be small, depending on $\beta$, such that $C\mu < \frac{1}{3}\beta$ and $C\epsilon\Delta\tau < \frac{1}{3}$. Then $\mathcal{F}_I$ maps $Q(\beta)$ into itself. One can also verify that if $\beta$ is small, then $\mathcal{F}_I$ is a contraction mapping. Therefore, there exists $\beta_0 > 0$ such that $\mathcal{F}_I : Q(\beta) \to Q(\beta)$ has a unique fixed point $\{u^i\}$.

Finally, the solution $\{u^i\}$ does not depend on the method of extending the domain of $\{\delta^i\}$, $\{g^i\}$ to $\tau \in \mathbb{R}^+$. This can be verified by using (15). □

**Remark.** In many formal constructions, $d = \inf\{\xi^{i+1} - \xi^i\} \to \infty$ as $\epsilon \to 0$. Then the condition $C_1 = Ce^{-\tilde{\alpha}d} \leq 0.5$ is valid if $\epsilon$ is small. We do not need to choose large $\gamma = M$ to make $\tilde{\alpha}$ large.

## PART II. Global Existence of Multiple Waves

**6.** The multiple wave solutions constructed in Part I exist only for a short time $t \in [0, \Delta t]$, $\Delta t = \epsilon\Delta\tau$. If $\{u^i\}$ is not too large, using the output of the previous interval as the input of the next time interval, the process can be repeated to obtain solutions in $[j\Delta t, (j+1)\Delta t]$, $j = 1, 2, \cdots$ recursively. It is shown, in [13], that if a formal approximation is defined for $t \in \mathbb{R}^+$s, under certain conditions, it is possible to obtain global solutions for $t \in \mathbb{R}^+$. In the second part of this paper, we summarize the results in [13].

Although the method should work for some higher order parabolic systems, such as the phase field equations, [7], to simplify the matter, we will consider a second order system,

(17) $$\epsilon u_t = \epsilon^2 u_{xx} + f(u, x, \epsilon), \quad x \in \mathbb{R}, t \geq 0.$$

Assuming by the method of matched expansions, we have the formal series for the wave fronts,

$$\eta^\ell(t, \epsilon) = \sum_0^m \epsilon^j \eta_j^\ell(t), \quad \ell \in \mathbb{Z},$$

and formal series solutions in the $\ell$-th regular and singular layers,

$$u^{R\ell}(x, t, \epsilon) = \sum_0^m \epsilon^j u_j^{R\ell}(x, t),$$

$$u^{S\ell}(\xi, t, \epsilon) = \sum_0^m \epsilon^j u_j^{S\ell}(\xi, t).$$

Here "R" and "S" stand for regular and singular (internal) layers, and $\xi = (x - \eta^\ell(t, \epsilon))/\epsilon$.

For convenience, assume the the following **Periodicity Hypotheses**.

1. $f(u, x + x_p, \epsilon) = f(u, x, \epsilon)$;
2. $\eta^{\ell+\ell_p}(t, \epsilon) = \eta^\ell(t, \epsilon) + x_p$;
3. $u^{R(\ell+\ell_p)}(x, t, \epsilon) = u^{R\ell}(x - x_p, t, \epsilon)$;
4. $u^{S(\ell+\ell_p)}(\xi, t, \epsilon) = u^{S\ell}(\xi, t, \epsilon)$.

Here $x_p > 0$ and integer $\ell_p > 0$ are two constants. The periodicity hypotheses ensure that all the estimates obtained here are uniform with respect to layer index $\ell$. They do not play any other rolls and are not necessary.

Let $0 < \beta < 1$ and let the width of the internal layers be $2\epsilon^{\beta-1}$. Define

$$y^{2\ell}(t) = \eta^\ell(t, \epsilon) + \epsilon^\beta,$$
$$y^{2\ell-1}(t) = \eta^\ell(t, \epsilon) - \epsilon^\beta.$$

The family of curves $\Gamma^i = \{(x, t) : x = y^i(t)\}$ divides the domain into regions $\Sigma^i$, $i \in \mathbb{Z}$, where $\Sigma^i$ is between $\Gamma^{i-1}$ and $\Gamma^i$. A formal approximation can be obtained from the matched expansions,

$$w^i(x, t, \epsilon) = \begin{cases} u^{R\ell}(x, t, \epsilon), & i = 2\ell - 1, \\ u^{S\ell}(\frac{x - \eta^\ell(t, \epsilon)}{\epsilon}, t, \epsilon), & i = 2\ell. \end{cases}$$

Here are the assumptions on $w^i$:

**H 1.** There exist $C, \bar{\gamma} > 0$ such that for all small $\epsilon$ and $i, \ell \in \mathbb{Z}$,

$$|w^i(x, t, \epsilon) - w^i(x, \infty, \epsilon)| \leq Ce^{-\bar{\gamma}t}, \quad (x, t) \in \mathbb{R}^2.$$
$$|\eta^\ell(t, \epsilon) - \eta^\ell(\infty, \epsilon)| + |D_t \eta^\ell(t, \epsilon)| \leq Ce^{-\bar{\gamma}t}, \quad t \in \mathbb{R}^+.$$

Here $w^i(x, \infty, \epsilon) = \lim_{t \to \infty} w^i(x, t, \epsilon)$, etc.

**H 2.** There exists $\bar{\sigma} > 0$ such that in each regular layer $\Sigma^i$, $i = 2\ell - 1$,

$$\text{Re}\,\sigma\{f_u(w^i(x,t,\epsilon), x, \epsilon)\} \leq -\bar{\sigma}$$

uniformly for all $(x,t) \in \Sigma^i$, $i \in \mathbb{Z}$ and small $\epsilon > 0$.

**H 3.** For an approximation $w^i(\xi, t, \epsilon)$ in an internal layer, as $\xi \to \pm\infty$ and $\epsilon \to 0$, both $w^i$ and $\partial w^i/\partial \xi$ approach the corresponding values of $w^{i+1}$ or $w^{i-1}$ at common boundaries. More precisely, if $i = 2\ell$, then for any $\mu > 0$, there exist $N$, $\epsilon_0 > 0$ such that $\epsilon_0^{\beta-1} > N$, and for $0 < \epsilon < \epsilon_0$, $t \geq 0$,

$$|W^i(\xi,t,\epsilon) - W^{i-1}(y^{i-1}(t),t,\epsilon)| \leq \mu, \quad \text{for } -\epsilon^{\beta-1} \leq \xi \leq -N,$$
$$|W^i(\xi,t,\epsilon) - W^{i+1}(y^i(t),t,\epsilon)| \leq \mu, \quad \text{for } \epsilon^{\beta-1} \geq \xi \geq N.$$

Here the function $W^i = (w^i, w^i_\xi)$ is expressed in the stretched variable $\xi = (x - \eta^\ell(t,\epsilon))/\epsilon$.

Let $\tilde{\xi} = \tilde{\xi}(\xi)$ be a function of $\xi$ such that

$$\tilde{\xi} = \begin{cases} \xi, & \text{for } |\xi| \leq \epsilon^{\beta-1}, \\ -\epsilon^{\beta-1}, & \text{for } \xi < -\epsilon^{\beta-1}, \\ \epsilon^{\beta-1}, & \text{for } \xi > \epsilon^{\beta-1}. \end{cases}$$

For each $t \geq 0$, $i = 2\ell$, consider the operator $\mathcal{A}^i(t) : L^2(\mathbb{R}) \to L^2(\mathbb{R})$,

$$\mathcal{A}^i(t)u = u_{\xi\xi} + D_t\eta^\ell(t,\epsilon)u_\xi + f_u(w^i(\tilde{\xi},t,\epsilon), \eta^\ell(t,\epsilon) + \epsilon\tilde{\xi}, \epsilon)u.$$

**H 4.** $\mathcal{A}^i(t)$, $i = 2\ell$, $t \leq 0$, has a simple eigenvalue $\lambda^i(\epsilon) = \epsilon\lambda_0^i(t) + O(\epsilon^2)$. The rest of the spectrum is contained in $\{\text{Re}\,\lambda \leq -\bar{\sigma}\}$, $\bar{\sigma}$ as in **H2**. Moreover, for the limiting operator $\mathcal{A}^i(\infty)$, we have,

$$\lambda_0^i(\infty) \leq \overline{\lambda}_0 < 0, \quad \text{uniformly for all } i = 2\ell.$$

We look for exact solution $u$ that is of the form $w^i + u^i$ in each $\Sigma^i$. The limit $u^i(x,\infty,\epsilon)$ describes the correction to $w^i(x,\infty,\epsilon)$ that yields a stationary solution, while $u^i(x,t,\epsilon) - u^i(x,\infty,\epsilon)$, together with $w^i(x,t,\epsilon) - w^i(x,\infty,\epsilon)$, describes how the solution approaches its limit as $t \to \infty$. Therefore, we define the following Banach spaces. For a constant $\gamma < 0$, let

$X(\Omega \times \mathbb{R}^+, \gamma) = \{u : u = u_1 + u_2,\ u_1 \in L^2(\Omega),\ u_2 \in L^2(\Omega \times \mathbb{R}^+, \gamma)\}.$
$|u|_{X(\gamma)} = |u_1|_{L^2(\Omega)} + |u_2|_{L^2(\Omega \times \mathbb{R}^+, \gamma)}.$
$X^{2,1}(\Omega \times \mathbb{R}^+, \gamma) = \{u : u = u_1 + u_2,\ u_1 \in H^2(\Omega),\ u_2 \in H^{2,1}(\Omega \times \mathbb{R}^+, \gamma)\}.$
$|u|_{X^{2,1}(\gamma)} = |u_1|_{H^2(\Omega)} + |u_2|_{H^{2,1}(\Omega \times \mathbb{R}^+, \gamma)}.$

It can be verified that for $u \in X(\Omega \times \mathbb{R}^+, \gamma)$ or $X^{2,1}(\Omega \times \mathbb{R}^+, \gamma)$, the decomposition $u = u_1 + u_2$ is unique.

For $\Sigma^i = \{(x,t) : y^{i-1}(t) < x < y^i(t),\ t \geq 0\}$, we say that $u \in L^2(\Sigma^i, \gamma)$, $H^{2,1}(\Sigma^i, \gamma)$, $X(\Sigma^i, \gamma)$ or $X^{2,1}(\Sigma^i, \gamma)$, etc., if $u$ is the restriction of a function $\tilde{u} \in$

$L^2(\mathbb{R} \times \mathbb{R}^+, \gamma)$, etc., to the domain $\Sigma^i$. The norms are defined by

$$|u|_{L^2(\Sigma^i,\gamma)} = \inf\{|\tilde{u}|_{L^2(\mathbb{R}\times\mathbb{R}^+,\gamma)}\},$$
$$|u|_{X^{2,1}(\Sigma^i,\gamma)} = \inf\{|\tilde{u}|_{X^{2,1}(\mathbb{R}\times\mathbb{R}^+,\gamma)}\}.$$

Let

$$X^k(\gamma) = \{u \mid u = u_1 + u_2,\ u_1 \in \mathbb{R}^n,\ u_2 \in H^k(\gamma)\},\ \gamma < 0.$$
$$X^{k_1 \times k_2}(\gamma) = X^{k_1}(\gamma) \times X^{k_2}(\gamma), \quad k_1 \geq 0,\ k_2 \geq 0.$$

Similar definitions can be given to spaces $L^2(\Omega \times I, \gamma)$, $H^{2,1}(\Omega \times I, \gamma)$, $H^{2,1}(\Sigma \cap I, \gamma)$, but not $X(\Omega \times I, \gamma)$ or $X^k(\gamma)$, if $I$ is a finite time interval.

The main result of Part II is the following *Global Spatial Shadowing Lemma*.

**Theorem 3.** *Let $\{\bar{w}^i\}$ be a formal approximation of solutions for (17). Assume that the Hypotheses* **H1**–**H4** *are satisfied, and the constants $\bar{\gamma}$ and $\bar{\sigma}$ satisfy $-\bar{\sigma} < -\bar{\gamma} < 0$. Then there exist positive constants $j_0$, $J_2$, $\epsilon_0$ and a negative constant $\gamma = O(\epsilon)$ satisfying the following properties. Assume that $\{w^i\}$ is a formal approximation near $\{\bar{w}^i\}$, with*

(18) $$|w^i - \bar{w}^i|_{X^{2,1}(\Sigma^i,\gamma)} \leq C_1 \epsilon^{j_1}, \quad i \in \mathbb{Z},$$

*for some $C_1 > 0$ and $j_1 \geq 1$. Assume that for the approximation $\{w^i\}$, we have*

(19) $$|g^i|_{X(\Sigma^i,\gamma)} + |\delta^i|_{X^{0.75 \times 0.25}(\mathbb{R}^+,\gamma)} \leq C_2 \epsilon^{j_2}, \quad j_2 \geq j_0.$$

*Then for $0 < \epsilon < \epsilon_0$, to any locally $H^1$ function $u_0$ with $|u_0 - w^i(0)|_{H^1(\Sigma^i \cap \{\tau=0\})} \leq C_2 \epsilon^{j_2}$, there exists a unique exact solution to (17) that satisfies $u_{exact}(x, 0) = u_0(x)$, and*

$$|u_{exact} - \bar{w}^i|_{X^{2,1}(\Sigma^i,\gamma))} = O(\epsilon^{j_3}), \quad i \in \mathbb{Z}.$$

*Here $j_3 = \min\{j_1, j_2 - J_2\}$, $J_2 > 0$ is a constant that does not depends on $\epsilon$. All the norms in this lemma are expressed by the stretched variables $\xi = x/\epsilon$, $\tau = t/\epsilon$.*

**Remark.** (1) The constants $j_0$ and $J_2$ depend on $\{\bar{w}^i\}$. However the approximation may not satisfy (19). By adding higher order expansions, the new approximation $\{w^i\}$ will satisfy (18). Therefore, it has the same constants $j_0$ and $J_2$, which are stable with respect to perturbations. Moreover, if the order of expansion is sufficiently high, $\{w^i\}$ will also satisfy (19).

(2) The values of $j_0$ and $J_2$ are not important in applications. since $j_3 = j_1$ if $j_2$ is sufficiently large. Computing $j_1$ is relatively easy. One only need to compare $\{w^i\}$ with the next approximation.

(3) Hypotheses **H1**–**H4** are consequences of hypotheses used to construct formal expansions for system (17). See §11 for a brief discussion.

**7.** To prove the theorem, we use a partition of the time interval $t \in \mathbb{R}^+$,

$$\mathbb{R}^+ = \cup_{j=0}^{r-1} [j\Delta t, (j+1)\Delta t] \cup [t_f, \infty),$$

where $t_f = r\Delta t$. We assume that $\Delta t = \epsilon \Delta \tau$, where $\Delta \tau$ is independent of $\epsilon$. $t_f$ is large such that $e^{-\bar{\gamma}t} = O(\epsilon^2)$, where $\bar{\gamma}$ is as in **H1**. If $\epsilon$ is small the variations of boundary curves $x = y^{i-1}(t)$ and $y^i(t)$ are small such that a near identity change of coordinates can be made in each $\Sigma^i_j = \Sigma^i \cap \{t \in [j\Delta t, (j+1)\Delta t]\}$, $j \leq r - 1$ or $\Sigma_r^i = \Sigma^i \cap \{t \in [t_f, \infty)\}$ to straighten $\Gamma^{i-1}$ and $\Gamma^i$. In the sequel, we assume that $\Gamma^i = \{x = y_j^i\}$ in the $j$th time interval, where $y_j^i$ is independent of $t$.

Define
$$\xi = [x - (y_j^{i-1} + y_j^i)/2]/\epsilon, \quad \text{in } \Sigma_j^i,$$
$$L_j^i(\epsilon) = (y_j^i - y_j^{i-1})/(2\epsilon),$$
$$\Omega_j^i = [-L_j^i(\epsilon), L_j^i(\epsilon)],$$
$$I^j = [0, \Delta\tau],\ 0 \le j \le r-1, \quad \text{and} \quad I^r = [0, \infty).$$

Let $\tau = (t - j\Delta t)/\epsilon$. Then $\Sigma_j^i = \Omega_j^i \times I^j$. In each $\Sigma_j^i$, $j \le r-1$, we linearize around $w^i$ at the fixed time $\tau = 0$, while in $\Sigma_r^i$, linearize around $w^i$ at the fixed time $\tau = \infty$. Denoting the solutions in $\Sigma_j^i$ by $u_j^i$, we have

(20) $\quad u_{j\tau}^i = u_{j\xi\xi}^i + V_j^i(\xi)u_{j\xi}^i + A_j^i(\xi)u_j^i + \mathcal{N}_j^i(u_j^i, \xi, \tau, \epsilon), \quad \text{in } \Sigma_j^i,$

(21) $\quad U_j^i(L^i(\epsilon), \tau) - U_j^{i+1}(-L^{i+1}(\epsilon), \tau) = \delta_j^i(\tau), \quad \text{at } \Gamma^i,$

(22) $\quad u_{j+1}^i(\xi, 0) = u_j^i(\xi_1, \Delta\tau),\ u_0^i(\xi, 0) = u_0^i(\xi).$

Here $\xi_1 = \Theta_j^i(\xi)$ is a near identity change of coordinates induced by the coordinates change that straightens the boundaries in each $\Sigma_j^i$. $V_j^i$ is the constant wave speed in an internal layer, but is slowly $\xi$ dependent in a regular layer due to the fact that the boundaries are not parallel there.

The idea of the proof is presented in the next three sections. In §8, we study the spectral properties of the linear systems and exponential dichotomies associated to the dual systems of the linear systems. In §9, we study the system in the last interval $[t_f, \infty)$. There we need to obtain asymptotic behavior of the solution $\{u_r^i\}$. Thus the space $X^{2,1}(\Sigma_r^i, \gamma)$ will be used. The result obtained there also serves as an upper bound of the accumulation error in the first $r$ intervals $[j\Delta t, (j+1)\Delta t]$, $0 \le j \le r-1$. In §10, we study the system in finite time intervals. Since $r$ is large, the estimates obtained in Part I is not accurate enough. More precise estimates will be derived which rely on the Hypotheses **H1**–**H4**, while the estimate in Part I does not.

**8.** Since nonlinear systems can be solved by contraction mappings, we study a linear system first. Drooping the subscript $j$, we consider,

(23) $\quad u_\tau^i = u_{\xi\xi}^i + V^i(\xi)u_\xi^i + A^i(\xi)u^i + h^i(\xi, \tau),$

(24) $\quad U^i(L^i(\epsilon), \tau) - U^{i+1}(-L^{i+1}(\epsilon), \tau) = \delta^i(\tau),$

(25) $\quad u^i(\xi, 0) = u_0^i(\xi).$

The homogeneous dual system of (23) is

(26) $$\hat{U}_\xi = \begin{pmatrix} 0 & I \\ sI - A^i(\xi) & -V^i(\xi) \end{pmatrix} \hat{U}.$$

Here $\hat{U} = (\hat{u}, \hat{u}_\xi)^\tau$. It is crucial to study the exponential dichotomies for system (26).

We first discuss the system in regular layers. Freezing $\xi = \xi_0$ and consider (26) with constant coefficients. Using **H2**, we can verify that for $\text{Re}(s) \geq -\sigma_0$, $\sigma_0 = \bar{\sigma}/2$, the system is hyperbolic, having $n$ eigenvalues with positive (negative) real parts.

Now recall that in regular layers, $A^i(\xi) = f_u(w^i((y^{i-1} + y^i)/2 + \epsilon\xi), (y^{i-1} + y^i)/2 + \epsilon\xi, \epsilon)$ depends slowly in $\xi$, i.e., $\frac{\partial A^i(\xi)}{\partial \xi} = O(\epsilon)$. Also, from the change of coordinates, cf. [13] for details, $V^i(\xi)$ also depends slowly on $\epsilon$ in regular layers. Then a theorem from [3] indicates that (26) has exponential dichotomy for $\xi \in \mathbb{R}$ and $\text{Re}(s) \geq -\sigma_0$.

In [13] it is shown that if (26) has an exponential dichotomy in $R^{2n}$ for $\xi \in \mathbb{R}$ and for every $s \in \{\text{Re}(s) \geq -\sigma_0\}$, then it has an exponential dichotomy in the space $E^{1,\nu}(s), \nu = 0$ or $0.25$.

Let the right hand side of (23) define a linear operator $\mathcal{A}^i(t)$, $t = j\Delta t$, $0 \leq j \leq r - 1$, or $t = \infty$, $j = r$. Using the fact that (26) has an exponential dichotomy in $E^{1,\nu}(s), \nu = 0$ for $\xi \in \mathbb{R}$ and $s \in \{\text{Re}(s) \geq -\sigma_0\}$, we can show that in regular layers, $\mathcal{A}^i(t)$ is an exponentially stable sectorial operator in $L^2(\mathbb{R})$, i.e.,

$$\text{Re}\{\sigma\mathcal{A}^i(t)\} \leq -\sigma_0,$$

for every $t \geq 0$.

We now discuss (26) in internal layers. Here the spectrum of $\mathcal{A}^i(t)$ is given by **H4**. Also $D_t\eta^\ell(t,\epsilon) = V^i_j$, $i = 2\ell$ if $t = j\Delta t$, $0 \leq j \leq r - 1$, and $V^i_r = 0$. From **H3**, if $\mu$ is small, then for $|\xi| \geq N$, the coefficients $A^i(\xi)$ and $V^i(\xi)$ are close to that of $A^i(\pm L^i(\epsilon))$ and $V^i(\pm L^i(\epsilon))$. But the system with constant coefficients $A^i(\pm L^i(\epsilon))$ and $V^i(\pm L^i(\epsilon))$ is hyperbolic. Therefore, from the Roughness of Exponential Dichotomies, which is also valid in spaces $E^{1,\nu}(s), \nu = 0, 0.25$, (26) has exponential dichotomies in $E^{1,\nu}(s), \nu = 0, 0.25$, for $\text{Re}(s) \geq -\sigma_0$ and $|\xi| \geq N$.

The exponential dichotomies in the region $\xi \leq -N$ and $\xi \geq N$ extend to $\xi \in \mathbb{R}^-$ and $\xi \in \mathbb{R}^+$ respectively. Let the projections be $P^i_s + P^i_u = I$. We can show that

$$(27) \quad \mathcal{R}P_u(0^-, s) \oplus \mathcal{R}P_s(0^+, s) = \mathbb{R}^{2n}, \quad \text{if } s \neq \lambda^i(\epsilon) \text{ and } \text{Re}(s) \geq -\sigma_0.$$

If (27) were not true, we could find a solution $\hat{U}(\xi, s)$ for (26) that decays exponentially as $\xi \to \pm\infty$. Then $s$ would be an eigenvalue with $\hat{U}(\xi, s)$ as an eigen function. This contradicts to the fact $\lambda^i(\epsilon)$ is the only eigenvalue in the region $\text{Re}(s) \geq -\sigma_0$.

In [13], it is also shown that the projections defined by the splitting (27) is $O(1 + \frac{1}{|s-\lambda^i(\epsilon)|})$. Observe that **H4** also implies that for the linearization at $t = \infty$, (26) has an exponential dichotomy in $\xi \in \mathbb{R}$ in the region $\text{Re}(s) \geq \frac{\epsilon\bar{\lambda}_0}{2}$, where $\bar{\lambda}_0 < 0$ is as in **H4**.

**9.** We now study (23)–(25) in the time interval $[t_f, \infty)$. The procedure of solving them is similar to that used in Part I. The right hand side of (23) defines a sectorial operator $\mathcal{A}^i(\infty)$ in $L^2(\mathbb{R})$. Extend the domain of $u_0^i$ and $h^i$ to $\xi \in \mathbb{R}$. Let

$$\bar{u}^i = e^{\mathcal{A}^i(\infty)\tau} u_0^i + \int_0^\tau e^{\mathcal{A}^i(\infty)(\tau-\zeta)} h^i(\zeta) d\zeta.$$

As shown in §8,

$$\sigma \mathcal{A}^i(\infty) \subset \{\text{Re}(s) \leq -C\epsilon < 0\},$$

both in regular and internal layers. Because of the spectra of $\mathcal{A}^i(\infty)$, with some $\gamma = C\epsilon < 0$, we can show

(28) $$|\bar{u}^i|_{X^{2,1}(\gamma)} \leq C(\epsilon^{-0.5}|u_0^i|_{H^1} + \epsilon^{-1.5}|h^i|_{X(\gamma)}).$$

The solution $\{u^i\}$ for (23)–(25) can be written as $u^i = \bar{u}^i + v^i$, with $\{v^i\}$ satisfying

(29) $$v_\tau^i = v_{\xi\xi}^i + V^i(\xi) v_\xi^i + A^i(\xi) v^i,$$

(30) $$V^i(L^i(\epsilon), \tau) - V^{i+1}(-L^{i+1}(\epsilon), \tau) = \eta^i(\tau),$$

(31) $$v^i(\xi, 0) = 0.$$

Here $V = (v, v_\xi)^\tau$, $\eta^i$ includes $\delta^i$ and corrections from the jumps of $\{\bar{u}^i\}$ at boundaries. We have

$$|\eta^i|_{X^{0.75 \times 0.25}(\gamma)} \leq C(|\delta^i|_{X^{0.75 \times 0.25}(\gamma)} + \epsilon^{-0.5}|\{u_0^i\}|_{H^1} + \epsilon^{-1.5}|\{h^i\}|_{L^2(\Sigma^i, \gamma)}).$$

Since for the dual system of (29)–(31), the exponential dichotomies exists in the region $\text{Re}(s) \geq \gamma$, $\gamma < 0$, using the Laplace transform and the iterative scheme similar to that used in Part I, we can show that system (29)–(31) has a unique solution $\{v^i\}$, with

$$|v^i|_{X^{2,1}(\gamma)} \leq C|\{\eta^i\}|_{X^{0.75 \times 0.25}(\gamma)}.$$

Therefore, in the last interval, the solution $\{u^i\}$ satisfies

$$|u^i|_{X^{2,1}(\gamma)} \leq C(|\{\delta^i\}|_{X^{0.75 \times 0.25}(\gamma)} + \epsilon^{-0.5}|\{u_0^i\}|_{H^1} + \epsilon^{-1.5}|\{h^i\}|_{L^2(\Sigma^i, \gamma)}).$$

Using the above result, the nonlinear system can be solved by a contraction mapping in a ball of radius $\epsilon^{r_1}$, $r_1 > 1.5$, in $X^{2,1}(\gamma)$. The following result is proved in [13].

**Theorem 4.** *Let $|g^i|_{X(\gamma)} = o(\epsilon^{r_1+1.5})$, $|u_0^i|_{H^1} = o(\epsilon^{r_1+0.5})$, $|\delta^i|_{H^{0.75 \times 0.25}(\gamma)} = o(\epsilon^{r_1})$, where $r_1 > 0.5$. Let $t_f$ be $\epsilon$ dependent such that $e^{-\bar{\gamma} t_f} \leq C_0 \epsilon^2$, where $\bar{\gamma}$ is as in **H1** and $C_0$ is independent of $\epsilon$. Then there exists $\epsilon_0 > 0$ such that for $0 < \epsilon < \epsilon_0$, the nonlinear system (20)–(22) has a unique solution $\{u^i\}$ satisfying the following estimate,*

(32) $$|\{u^i\}|_{X^{2,1}(\gamma)} \leq C\{|\{\delta^i\}|_{H^{0.75 \times 0.25}(\gamma)} + \epsilon^{-0.5}|\{u_0^i\}|_{H^1} + \epsilon^{-1.5}|\{g^i\}|_{X(\gamma)}\}.$$

**10.** We study system (20)–(22) in the finite intervals, $[j\Delta t, (j+1)\Delta t]$, $0 \leq j \leq r-1$. The existence of solutions in a finite interval has been discussed in Theorem 2. However, to guarantee the existence of solutions up to the last interval $[t_f, \infty)$, a stringent restriction on the accumulation errors in these finite intervals must be met. Since the critical eigenvalue $\lambda^i(\epsilon)$ may not be negative, $u_j^i$ may grow as $j$ increases. From Theorem 2, $\{u_j^i\}$ is determined by $\{u_j^i(0)\}$, $\{g_j^i\}$ and $\{\delta_j^i\}$. Among them only $\{u_j^i(0)\}$, carries the information from the previous $j$ intervals. Therefore, it is crucial to control $|u_j^i(\Delta\tau)|$ in terms of $|\{u_j^i(0)\}|$.

Let $Q_0^{i,j}$ and $Q_s^{i,j}$ be the projections corresponding to the spectral set in $\lambda^i(\epsilon)$ and $\{\text{Re}(s) \leq -\sigma_0\}$ in the $i$th internal layers and the $j$th time interval. Let $Q_0^{i,j} u_j^i = \alpha_j^i \phi_j^i$ where $\phi_j^i$ is the eigenvalue corresponding to the eigenvalue $\lambda^i(\epsilon)$. Then
$$|\alpha_j^i| + |Q_s^{i,j} u_j^i|_{H^1},$$
is an equivalent norm for $|u_j^i|_{H^1}$. The following estimate has been proved in [13],

$$\begin{aligned}(33)\quad |\alpha_j^i(\Delta\tau)| + |Q_s^{i,j} u_j^i(\Delta\tau)|_{H^1} &\leq (1+C\epsilon)\sup_i(|\alpha_j^i(0)| + |Q_s^{i,j} u_j^i(0)|_{H^1}) \\ &\quad + C|\{\delta_j^i\}|_{H^{0.75\times 0.25}(I^j)} + C|\{g_j^i\}|_{L^2}.\end{aligned}$$

Since the spectral projections for the $j$ and $(j+1)$th interval differ by $O(\epsilon)$, $|\alpha_{j+1}^i(0)| + |Q_s^{i,j+1} u_{j+1}^i(0)|_{H^1}$ is also bounded by the right hand side of (33). Therefore, one can show that

$$|\alpha_j^i(0)| + |Q_s^{i,j} u_j^i(0)|_{H^1} \leq (1+C\epsilon)^j \sup_i(|\alpha_0^i(0)| + |Q_s^{i,0} u_0^i|_{H^1})$$
$$+ \frac{(1+C\epsilon)^j}{\epsilon} \sup_{k<j}(|\{\delta_k^i\}|_{H^{0.75\times 0.25}} + |\{g_k^i\}|_{L^2}).$$

Using the fact $(1+C\epsilon)^{1/(C\epsilon)} < e$ and $j \leq r \approx \log(\frac{1}{\epsilon})/\epsilon$, we see that $(1+C\epsilon)^j \leq \epsilon^{-B}$ for some $B > 0$. Therefore, if all the terms in the right hand side are bounded by $\epsilon^M$ for some large $M > 0$, have the solution $\{u_j^i\}$ for all the finite intervals, and the initial condition for the infinite interval is small enough so that Theorem 4 can be used there. Here is the main theorem in this section.

**Theorem 5.** *There exists $\Delta\tau > 0$ such that if $r = [\log(\frac{1}{C_0\epsilon})/\epsilon\bar{\gamma}\Delta\tau] + 1)$, then we have the following concerning the solutions of (20)–(22) in $I^j$, $0 \leq j \leq r-1$. There exist $M_1$ and $M_2 > 0$, such that if*
$$|\{u_0^i(0)\}|_{H^1} + |\{\delta_j^i\}|_{H^{0.75\times 0.25}} + |\{g_j^i\}|_{L^2} = O(\epsilon^{M_1}),$$
*uniformly for $0 \leq j \leq r-1$, then (20)–(22), $0 \leq j \leq r-1$ has a unique solution $u_j^i$ for $0 \leq \epsilon \leq \epsilon_0$, where $\epsilon_0 > 0$ is a small constant. The solution in each $I^j$ satisfies*
$$|\alpha_j^i(0)| + |Q_s^{i,j} u_j^i(0)|_{H^1} \leq C\epsilon^{-M_2}[\sup_i(|\alpha_0^i(0)| + |Q_s^{i,0} u_0^i|_{H^1})$$
$$+ \sup_{k<j}(|\{\delta_k^i\}|_{H^{0.75\times 0.25}} + |\{g_k^i\}|_{L^2}].$$
*In particular, if $M_1$ is sufficiently large, in the rth (infinite) interval $|\{u_r^i(0)\}|_{H^1} = o(\epsilon^{r_1+0.5})$, $r_1 > 1.5$. Therefore, Theorem 4 applies in the last infinite interval also.*

To describe the idea of the proof of (33), it suffices to consider the linear system (23)–(25). For simplicity, we consider the 0th interval and drop the subscript $j$. The procedure of solving such system is similar to that used before. Define $\bar{u}^i$ as in Part I, and let $\bar{u}^i(\tau) = \bar{\alpha}^i(\tau)\phi^i + Q_s^{i,0}\bar{u}^i(\tau)$. In internal layers, $\lambda^i = O(\epsilon)$. Also in the stable space, the semigroup is exponentially stable. We can show,

$$|\bar{\alpha}^i(\Delta\tau)| \le e^{(1+C\epsilon)\Delta\tau}(|\alpha^i(0)| + \sqrt{\Delta\tau}|h^i|_{L^2}),$$
$$|Q_s^{i,0}\bar{u}^i(\Delta\tau)|_{H^1} \le C(e^{-\sigma\Delta\tau}|Q_s^{i,0}u^i(0)|_{H^1} + |\alpha^i(0)|),$$

for some $\sigma < 0$. Observe in the above, $\bar{u}^i(0) = u^i(0)$. Thus $\bar{\alpha}^i(0) = \alpha^i(0)$.

Let $\Delta\tau > 0$ be large so that $Ce^{-\sigma\Delta\tau} \le 1/2$. We then find $C_1$ such that $e^{(1+C\epsilon)\Delta\tau} \le 1 + C_1\epsilon$. We have

$$(34) \quad |\bar{\alpha}^i(\Delta\tau)| + |Q_s^{i,0}\bar{u}^i(0)|_{H^1} \le (1+C_1\epsilon)|\alpha^i(0)| + 1/2|Q_s^{i,0}u^i(0)|_{H^1} + C|h^i|_{L^2}.$$

The above is also valid in regular layers with $\bar{\alpha}^i = 0$.

The solution for (23)–(25) again is written as $u^i = \bar{u}^i + v^i$, where $\{v^i\}$ satisfies (29)–(31). As in Part I, system (29)–(31) is solved by iterations, and $\{v^i\}$ is bounded by $\{\eta^i\}$ in suitable function spaces. An complication arises here since $\{\eta^i\}$ also include corrections of the jumps of $\{\bar{u}^i\}$ at common boundaries. The latter in term is bounded by $|\{\alpha^i(0)\}|$ and $|\{Q_s^{i,0}u^i(0)\}|$, of which the coefficients have to be carefully controlled.

It can be shown that $\{\eta^i\}$ depends weakly on $\alpha^i(0)$, since $|\alpha^i\phi^i| \le Ce^{-\sigma|\xi|}$ for some $\sigma > 0$, and $L^i(\epsilon) \to \infty$ as $\epsilon \to 0$. However, it is a little tricky to control the dependence of $|\{\eta^i\}|$ on $|\{Q_s^{i,0}u^i(0)\}|$. The key here is to use the exponential decay of $e^{\mathcal{A}^i\tau}$ in the stable subspace (which is the whole space in regular layers). To do so, weighted norms have been used, [13],

$$|U|_{H^{0.75\times 0.25}(I^j,\sigma)} = |e^{-\sigma\tau}U|_{H^{0.75\times 0.25}(I^j)}, \quad \sigma < 0,$$
$$|u|_{H^{2,1}(\Omega^i\times I^j,\sigma)} = |e^{-\sigma\tau}u|_{H^{2,1}(\Omega^i\times I^j)}, \quad \sigma < 0.$$

Observe here the interval $I^j = [0,\Delta\tau]$ is finite. It is shown that

$$|\eta^i|_{H^{0.75\times 0.25}(I^j,\sigma)} \le Ce^{|\sigma|\Delta\tau}|\{\delta^i\}|_{H^{0.75\times 0.25}(I^j)}$$
$$+ \sup_i\{\epsilon|\alpha^i(0)| + C|Q_s^{i,0}u^i(0)| + Ce^{|\sigma|\Delta\tau}|h^i|_{L^2}\},$$

for some $\sigma < 0$, independent of $\epsilon$. Also it is shown that the solution of the system (29)–(31) satisfies

$$|v^i|_{H^{2,1}(\Omega^i\times I^j,\sigma)} \le C|\{\eta^i\}|_{H^{0.75\times 0.25}(I^j,\sigma)}.$$

Let $Q_0^{i,0}v^i = \bar{\bar{\alpha}}^i\phi^i$.

$$|\bar{\bar{\alpha}}^i(\Delta\tau)| + |Q_s^{i,0}v^i(\Delta\tau)|_{H^1} \le C|v^i(\Delta\tau)|_{H^1}$$
$$\le Ce^{\sigma\Delta\tau}|v^i|_{H^{2,1}(\Omega^i\times I^j,\sigma)}$$
$$\le C|\{\delta^i\}|_{H^{0.75\times 0.25}(I^j)}$$
$$+ \sup_i\{C_1e^{\sigma\Delta\tau}|Q_s^{i,j}u^i(0)|_{H^1} + C\epsilon e^{\sigma\Delta\tau}|\alpha^i(0)| + C|h^i|_{L^2}\}.$$

Let $\Delta\tau$ be large so that $C_1 e^{\sigma \Delta\tau} |Q_s^{i,j} u^i(0)|_{H^1} \leq 0.5$. Combine the above with (34), we have an estimate for the linear system,

$$
\begin{aligned}
|\alpha^i(\Delta\tau)| + |Q_s^{i,0} u^i(\Delta\tau)|_{H^1} &\leq (1+C\epsilon) \sup_i (|\alpha^i(0)| + |Q_s^{i,0} u^i(0)|_{H^1}) \\
&\quad + C|\{\delta^i\}|_{H^{0.75 \times 0.25}(I^j)} + C|\{h^i\}|_{L^2}.
\end{aligned}
\tag{35}
$$

The estimate of the nonlinear system (33) can be obtained from (35) by replacing $h^i$ with $\mathcal{N}^i$, and using the fact that if $|u^i|_{H^{2,1}} < \epsilon$,

$$|\mathcal{N}^i|_{L^2} \leq |g^i|_{L^2} + C\epsilon |u^i|_{H^{2,1}}.$$

**11.** Constructions of multiple waves in singularly perturbed equations have been discussed in many papers, see for example [5]. Expansions for system of reaction-diffusion equations to any order of $\epsilon$ can be found in [12]. The following hypotheses are used in that paper.

There is a partition

$$\mathbb{R} = \cup_{\ell=-\infty}^{\infty} [x^\ell, x^{\ell+1}]$$

that is periodic with respect to $\ell$, compatible with the period of $f$. A $C^\infty$ function $p^i(x)$ is defined on $[x^{i-1}, x^i]$ with $f_0(p^i(x), x) = 0$. Also assume that $p^{\ell+\ell_p}(x+x_p) = p^\ell(x)$, $x \in [x^{\ell-1}, x^\ell]$.

**H\* 1** $\operatorname{Re}\sigma\{f_{0u}(p^i(x), x)\} < 0$ for $x \in [x^{i-1}, x^i]$, $i \in \mathbb{Z}$.

Using a stretched variable $\xi = \dfrac{x - x^i}{\epsilon}$, we assume that the 0th order expansion of Eq. (17),

$$u_{\xi\xi} + f_0(u, x^i) = 0. \tag{36}$$

has a heteroclinic solution $q^i(\xi)$ connecting $p^i(x^i)$ to $p^{i+1}(x^i)$. Assume

**H\* 2** The linear homogeneous equation

$$\phi_{\xi\xi} + f_{0u}(q^i(\xi), x^i)\phi = 0,$$

has a unique bounded solution $q_\xi^i(\xi)$, up to constant multiples.

From **H\*2**, we infer that the adjoint equation

$$\psi_{\xi\xi} + f_{0u}^\tau(q_\xi^i(\xi), x^i)\psi = 0$$

has a unique bounded solution $\psi_i(\xi)$ up to constant multiples.

The following assumption implies that the heteroclinic solution breaks as $x$ moves away from $x^i$.

**H\* 3** $\int_{-\infty}^{\infty} \psi_i^\tau(\xi) f_{0x}(q^i(\xi), x^i) d\xi \neq 0, \quad i \in \mathbb{Z}$.

When $x$ is in a neighborhood of $x^i$, we look for traveling wave solutions with wave speed $V^i(x)$. It is also of interest to find out conditions to ensure that the wave front moves towards $x^i$. Consider $\mathcal{A}^i : L^2(\mathbb{R}) \to L^2(\mathbb{R})$, defined as

$$\mathcal{A}^i u = u_{\xi\xi} + f_{0u}(q^i(\xi), x^i) u, \quad D(\mathcal{A}^i) = H^2(\mathbb{R}).$$

**H\* 4** $\lambda = 0$ is a simple eigenvalue for $\mathcal{A}^i$, $i \in \mathbb{Z}$. There exists $\sigma_0 > 0$ such that
$$\sigma(\mathcal{A}^i) \subset \{\lambda = 0\} \cup \{\text{Re}\lambda \leq -\sigma_0\}.$$

Since $\lambda = 0$ is simple, we have
$$\int_{-\infty}^{\infty} \psi_i^\tau(\xi) q_\xi^i(\xi) d\xi \neq 0.$$

**H\* 5** $I = [\int_{-\infty}^{\infty} \psi_i^\tau(\xi) q_\xi^i(\xi) d\xi]^{-1} \int_{-\infty}^{\infty} \psi_i^\tau(\xi) f_{0x}(q^i(\xi), x^i) d\xi > 0$, $i \in \mathbb{Z}$.

Under Hypotheses **H\*1**–**H\*5**, it is shown, in [12], that formal expansions of wave front positions $\{\eta^i(t, \epsilon)\}$ and matched expansions of solutions $\{w^i(x, t, \epsilon)\}$ can be constructed. These hypotheses are general in the sense that **H\*2** and **H\*3** are generic assumptions. Hypothesis **H\*1** is equivalent to the stability of $p^i(x)$ in a regular layer, as an equilibrium of an ODE, obtained from the 0th expansion of (17), i.e., by setting $\epsilon u_{xx} = 0$. Hypothesis **H\*4** is equivalent to the stability of the traveling wave solution $q^i$ in an internal layer, in the sense of Evans [4]. Hypothesis **H\*5** implies that the wave front is moving towards $x^i$, if $x$ is near $x^i$. In fact, if is shown in [12]
$$\frac{\partial V^i(x^i)}{\partial x} = -I \cdot (x - x^i).$$
Similar conditions like **H\*5** was used by Fife [5]. Without such conditions, the wave front may move away from the stationary position and the formal solutions may not exist globally in time.

It is shown in [13] that Hypotheses **H\*1**–**H\*5** imply Hypotheses **H1**–**H4** of this paper. Only **H4** needs some explanation. When $\{w^i\}$ is a 0th order approximation, $\lambda = 0$ is always an eigenvalue. When higher order terms are added, the eigenvalue becomes $O(\epsilon)$. In [13], a precise formula expressing $\lambda^i(\epsilon)$ as a function of the wave speed and the wave front position was proved. In particular, when $t = \infty$, it yields
$$\frac{\partial \lambda^i}{\partial \epsilon} = \frac{\partial V^i}{\partial x}.$$
Here $V^i x)$ is the wave speed as a function of the wave position $x$. It is now clear that **H\*5** implies **H4**.

Many authors have found that the signs of $\lambda^i(\epsilon)$ or the derivatives of the wave speed determine the stability of the multiple waves, [6]. The precise relation between the two quantities in general systems is interesting in its own right right.

In conclusion, the formal approximations obtained in [12] satisfies our hypotheses in Part II. Therefore, the *Global Spatial Shadowing Lemma* can be used to ensure the existence of a global solution near the formal approximation.

## References

1. N. Alikakos, P. Bates and G. Fusco, *Slow motion for the Cahn–Hilliard equation in one space dimension*, J. Differential Equations **90** (1991), 81–135.
2. S.-N. Chow, X.-B. Lin and K. Palmer, *A shadowing lemma with applications to semilinear parabolic equations*, SIAM J. Math. Anal., **20**, (1989), 547–557.

3. W. A. Coppel, *Dichotomies in stability theory*, A. Dold & B. Eckmann eds. Lecture Notes in Math. **629**, Spring-Verlag, New York, 1970.
4. John W. Evans, *Nerve Axon Equations: I Linear Approximations*, Indiana Univ. Math. J., **21** (1972), 877–885
5. P. C. Fife, *Pattern formation in reacting and diffusing systems*, J. Chem. Phys., **64** (1976), 554–564.
6. G. Fusco and J. Hale, *Slow motion manifolds, dormant instability, and singular perturbations*, J. Dynamics and Differential Equations **1** (1989), 75–94.
7. R. A. Gardner & C. K. R. T. Jones, *Traveling waves of a perturbed diffusion equation arising in a phase field model*, Indiana Univ. Math. J., **38** (1989), 1197-1222.
8. J. Guckenheimer, J. Moser & S. Newhouse, Dynamical Systems, Birkhäuser, Boston, 1980.
9. D. Henry, *Geometric theory of semilinear parabolic equations*, Lecture Notes in Math. vol **840**, Springer-verlag, 1982.
10. K. Kirchgässner, *Homoclinic bifurcation of perturbed reversible systems*, W. Knobloch and K. Smith, eds., Lecture Notes in Math., 1017 (1982), Springer-Verlag, New York, 328-363.
11. X.-B. Lin, *Shadowing lemma and singularly perturbed boundary value problems*, SIAM J. Appl. Math., **49** (1989), 26–54.
12. X.-B. Lin, *Asymptotic expansion for layer solutions of a singularly perturbed reaction-diffusion system*, preprint, 1994.
13. X.-B. Lin, *Shadowing matching errors for wave-front like solutions*, preprint
14. J. L. Lions and E. Magenes, *Non-homogeneous boundary value problems and applications I*, Springer-Verlag, New York, 1972.
15. M. Renardy, *Bifurcation of singular solutions in reversible systems and applications to reaction-diffusion equations*, Advances in Mathematics **3**, (1982), 324-406.
16. K. Yosida, *Functional Analysis*, Grundlehren der mathematischen Wissenschaften **123**, Springer, New York, 1980.

Department of Mathematics, North Carolina State University, Raleigh, North Carolina 27695–8205

*E-mail address*: `xblin@@xblsun.math.ncsu.edu`